\documentclass[aip,jcp,reprint]{revtex4-1}
\usepackage{amsmath}
\usepackage{latexsym}
\usepackage{float}
\usepackage{amssymb}
\usepackage{graphicx}
\usepackage{epsfig}
\usepackage{psfrag}
\usepackage{color}
\usepackage{multirow}
\begin{document}

\title{One size fits all: equilibrating chemically different polymer liquids through universal long-wavelength description}

\author{Guojie Zhang}
\affiliation{Max Planck Institute for Polymer Research, Ackermannweg 10, 55128 Mainz, Germany}

\author{Torsten Stuehn}
\affiliation{Max Planck Institute for Polymer Research, Ackermannweg 10, 55128 Mainz, Germany}

\author{Kostas Ch. Daoulas}
\affiliation{Max Planck Institute for Polymer Research, Ackermannweg 10, 55128 Mainz, Germany}

\author{Kurt Kremer}
\altaffiliation{Corresponding author e-mail: kremer@mpip-mainz.mpg.de}
\affiliation{Max Planck Institute for Polymer Research, Ackermannweg 10, 55128 Mainz, Germany}

\begin{abstract}
Mesoscale behavior of polymers is frequently described by universal laws. This physical property motivates us to propose a new modeling concept, grouping polymers 
into classes with a common long-wavelength representation. In the same class samples of different materials can be generated from this representation, 
encoded in a single library system. We focus on homopolymer melts, grouped according to the invariant degree of polymerization. 
They are described with a bead-spring model, varying chain stiffness and density to mimic chemical diversity. In a renormalization group-like fashion 
library samples provide a universal blob-based description, hierarchically backmapped to create configurations of other class-members. 
Thus large systems with experimentally-relevant invariant degree of polymerizations (so far accessible only on very coarse-grained level) can be microscopically described. 
Equilibration is verified comparing conformations and melt structure with smaller scale conventional simulations.
\end{abstract}

\maketitle

Predicting properties of polymeric materials with computer simulations often requires their description with microscopic detail. 
This is, however, challenging due to slow dynamics of entangled ``spaghetti-like'' polymers \cite{Doi} and the need to address systems with dimensions significantly larger than the 
coil size (to avoid finite-size effects). To circumvent these difficulties, hierarchical strategies gradually equilibrating the material on different observation 
scales are attractive (see ref.~\cite{PeterKremer} and references therein). First the crudest level is addressed with models, representing a large amount of 
microscopic degrees of freedom by a single effective particle. Details on shorter wavelengths are then gradually reinserted 
until all chemical details are recovered. Computations remain tractable since backmapping requires only local sampling.

Implicitly hierarchical approaches assume that the long-wavelength structure is correctly 
described by the crude model and is insensitive to chemical details. Scale-decoupling in polymers can be rigorously justified and linked to universality of the long-wavelength 
behavior, often described by generic laws adsorbing chemistry-specific details into few parameters \cite{deGennes}. 
Since different polymers can be mapped to the same point of parameter space, classes of systems with identical 
long-wavelength properties can be defined. This allows simplifying significantly the hierarchical 
modeling of polymeric materials by creating ``material-genomic'' libraries of morphologies with the correct
long-wavelength behavior of different classes. For an entire class such morphology must be generated only once, 
preferably using the most simple microscopic representation. This system can be even a chemistry non-specific model (e.g., bead-spring) which nevertheless maps 
on the same point of parameter space. Atomistic representations of any class-member can be recovered reinserting chemical details into the common long-wavelength description 
with standard techniques~\cite{PeterKremer}. 

Here the implementation of this concept is demonstrated for amorphous homopolymer melts. Despite their simplicity, 
they present significant interest for basic polymer physics (e.g., as a framework for studying rheology) and industry. 
Polymer melts are systems where long strongly interdigitating molecules fill space in a random walk fashion. The number of molecules crossing the volume of a test chain in a melt of 
type~\cite{typexplan} $\gamma$ is proportional to $\bar{N}_{\rm (\gamma)}^{1/2} = \rho_{\rm (\gamma)} R_{\rm (\gamma)}^3 / N_{\rm (\gamma)}$, 
where $\rho_{\rm (\gamma)}$, $R_{\rm (\gamma)}$, and $N_{\rm (\gamma)}$ are the monomer density, root mean-square end-to-end distance, and 
polymerization degree respectively. The invariant degree of polymerization, $\bar{N}_{\rm (\gamma)}$, presents a natural choice for arranging homopolymer melts 
into groups with common static long-wavelength properties. Indeed mesoscale conformations~\cite{Strasbourg, Morse2014MM} and liquid structure~\cite{deGennes, MuellerBinder, Guenza2013} are known to be universal 
functions of this quantity (see below). Generally for dynamical properties non-universalities can appear even in the long-wavelength limit~\cite{McLeish}, however there is evidence 
that the invariant degree of polymerization of subchains between two consecutive entanglements is similar for all polymers~\cite{Lin1987, Fetters1994, RalfScience}. 
$\bar{N}_{\rm (\gamma)}$ plays a key role in more complex systems, e.g. it controls in a universal way thermodynamic properties in symmetric block-copolymers~\cite{Fredrickson1987, Morse2014}.
\begin{figure}[ht]
\includegraphics[width=0.48\textwidth]{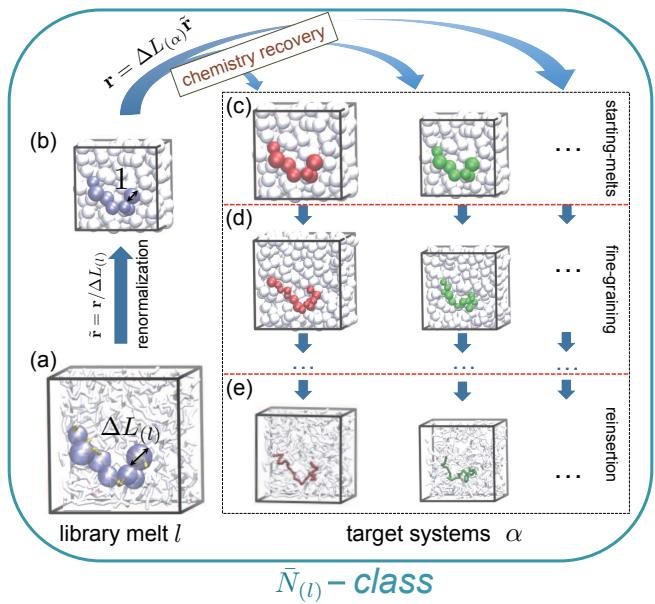}
\caption{Hierarchical modeling scheme for homopolymer melts with the same invariant degree of polymerization, ${\bar{N}_{(l)}}$, forming a single class 
of materials. (a) A library configuration described with microscopic detail is subjected to coarse-graining (the average blob size is $\Delta L_{(l)}$) 
and scaling of coordinate-space, $\tilde{\bf{r}} = {\bf r}/\Delta L_{(l)}$, to obtain (b) a universal blob-based description of long-wavelength structure (the average blob size is unity). 
(c) The universal description is projected on the coordinate space of any other target $\alpha$-type melt of the same class, back-transforming 
the coordinate space,  ${\bf r} = \Delta L_{\rm (\alpha)} \tilde{\bf{r}}$. The scale $\Delta L_{(\alpha)}$ must be properly chosen (see main text). 
(d) The initial blob-based description undergoes a sequence of fine-graining steps, substituting every blob in the preceding representation by a pair of smaller ones. 
(e) Once the blobs are sufficiently small, microscopic details can be reinserted.} 
\label{fig:summary}
\end{figure}

Let us consider different melts (indexed by $\alpha$), having the same ${\bar{N}_{(l)}}$ as the $l$-th library melt (see Fig.~\ref{fig:summary}). 
For each melt the regime of universal long-wavelength behavior can be defined introducing a coarse-graining length scale, 
$\Delta L_{\rm (\gamma)}$ ($\gamma = \alpha, l$), in spirit of renormalization group theories \cite{deGennes, Freed}. In principle the observation scale separating chemistry-specific 
and universal behavior cannot be exactly defined, thus $\Delta L_{\rm (\gamma)}$ presents at this stage an arbitrary ``large'' scale. 
Introducing $\Delta L_{\rm (\gamma)}$ renormalizes the microscopic coordinate space as $\tilde{\bf{r}} = {\bf r}/\Delta L_{\rm (\gamma)}$. 
The shortest subchains that are resolved during coarse-graining contain a number of monomers, $N_{b{\rm (\gamma)}}$, such 
that their root mean-square end-to-end distance, $R_{b{\rm (\gamma)}}$, equals $\Delta L_{\rm (\gamma)}$. $\Delta L_{\rm (\gamma)}$ 
can be simply related to $N_{b{\rm (\gamma)}}$ within the Flory hypothesis (FH)~\cite{Flory1949, deGennes}. Linking monomers into 
long chains in combination with incompressibility reduces intermolecular monomer-monomer contacts on the scale of chain-size creating 
a ``correlation hole''. FH proposes that intramolecular interactions tending to swell a chain are canceled by interactions with the other molecules, forming the ``soft walls'' 
of the correlation hole. Hence the polymer conformations on the mesoscale are ideal random walks, so that $R_{b{\rm (\gamma)}}^2 = N_{b{\rm (\gamma)}} b_{\rm e (\gamma)}^2$. 
Thus $N_{b{\rm (\gamma)}} = \Delta L_{\rm (\gamma)}^2/b_{\rm e (\gamma)}^2$. The chemistry-specific coefficient, $b_{\rm e (\gamma)}^2$, is the squared effective bond length~\cite{Doi}. 

All melts are represented in the renormalized space by ensembles of chains of spheres (blobs) with the same average 
diameter $\tilde{R}_{b{\rm (\gamma)}} = 1$. Combining known theoretical results, we argue that under certain conditions such an ensemble constitutes a common long wavelength 
description for melts at given ${\bar{N}_{(l)}}$. In practice, stored (library) configurations will have a fixed number of chains, $n_{(l)}$, and volume, $V_{(l)}$. 
Hence to identify these conditions it is natural to consider the canonical ensemble. The $\alpha$-type and the 
library melt will have the same conformations and structure in the renormalized space when (a) they are 
represented by the same amount of blobs, $N_{\rm CG} = {\bar{N}_{(l)}}/\rho_{\rm (\gamma)}^2 b_{\rm e (\gamma)}^4 \Delta L_{\rm (\gamma)}^2$~\cite{blobexplan}, 
(b) contain an equal number of chains, $n$, and (c) have the same rescaled volume, $\tilde{V} = V_{\rm (\gamma)}/\Delta L_{\rm (\gamma)}^3$ (in fact, 
since $\bar{N}_{\rm (\gamma)}$ is the same, the last condition follows from the other two). 

When ideal random walks are coarse-grained into a blob representation, the statistics: (a) of the distance, $\tilde{l}$, between the center-of-mass (COM) of two sequential blobs
and (b) of the angle, $\theta$, between two vectors joining the COM of a blob with the COMs of the preceding and the succeeding blob, are universal in renormalized space~\cite{Suter1991}.
Since the distributions of internal coordinates $\tilde{l}$ and $\theta$ are universal, the conformations of coarse-grained chains in the library and $\alpha$-type melts
with an identical amount of blobs will be the same. The average mean-square distance of two blobs with ranking numbers $s_{\rm 1}$ and $s_{\rm 2}$ along the
chain contour is a representative quantifier of these conformations, given by:
\begin{eqnarray}
\tilde{R}_{\rm CG}^2(s) = s \langle\tilde{l}^2\rangle \left[\frac{1 + p}{1 - p } - \frac{2 p (1-p^s)}{s(1-p)^2} \right]
\label{eqn:internaldist}
\end{eqnarray}
$\langle\tilde{l}^2\rangle = \frac{2}{3}$, $p=\langle\cos{(\pi - \theta)}\rangle\simeq 0.22$, and $s=|s_{\rm 2} - s_{\rm 1}|$.

FH was corrected~\cite{Strasbourg, Morse2014MM}, demonstrating that the volume spanned by a chain presents a hierarchy of nested correlation 
holes of all possible subchains. This leads to weak self-avoidance and deviations from the ideal random walk behavior. The power-law dependence 
is replaced by~\cite{Strasbourg, Morse2014MM}:
\begin{eqnarray}
R_{b{\rm (\gamma)}}^2 = 
N_{b{\rm (\gamma)}} b_{\rm e (\gamma)}^2 \left[1 - c_1{N_{b{(\gamma)}}}^{-1/2} + c_2 N_{b{(\gamma)}}^{-1}\right]
\label{eqn:strasbourg}
\end{eqnarray} 
$c_1$ and $c_2$ are constants which, in general, are chemistry-specific (though for the end-to-end distance of long chains, the leading 
order correction in Eq.~\ref{eqn:strasbourg} reduces~\cite{Morse2014} to a universal form).

Since the melts have the same number of chains it is straightforward to show that they have the same invariant degree of polymerization also on the level of subchains, 
where $\bar{N}_{b{\rm(\gamma)}}^{1/2} = n N_{CG}/ \tilde{V}$. Due to the same $\bar{N}_{b{\rm(\gamma)}}$ the liquid structure on the scale of blobs will be the same. 
This conclusion can be traced back to de Gennes predicting~\cite{deGennes} that the average depth of the correlation hole for entire chains 
scales as $\sim \bar{N}_{{\rm(\gamma)}}^{-3/2}$. Modeling studies highlighted the universality of the shape of the correlation 
hole on intermediate scales~\cite{MuellerBinder}. Blob-packing has been considered within integral equation theory predicting for their intermolecular pair distribution function~\cite{Guenza2013}: 
\begin{eqnarray}
g_{CG}(\tilde{r}) = 1 - A_{\rm o}\bar{N}_{b(\gamma)}^{-1/2} X_0(\tilde{r}, N_{CG})
\label{eqn:guenza}
\end{eqnarray}
where $A_{\rm o} = 6^{3/2}/2\pi^2$ and $X_0$ is a universal function. 

In summary, theoretical arguments suggest that configurations of homopolymer melts with the same invariant degree of polymerization 
can be obtained by backmapping a common blob-based representation. The latter is defined in spaces renormalized 
by the coarse-graining scale and originates from a library melt. In all practical applications, chain length is defined 
in terms of $N_{\rm (\alpha)}$ thus the matching library configuration is chosen such that, 
${\bar{N}_{(l)}}^{1/2} = \rho_{\rm (\alpha)} b_{\rm e (\alpha)}^3 N_{\rm (\alpha)}^{1/2}$. The coarse-graining scales of the 
modeled and library melts are arbitrary, provided that: (a) $\Delta L_{\rm (\gamma)}$ are sufficiently large for the subchains 
to be approximated by ideal random walks and (b) $\Delta L_{\rm (\alpha)} = (\rho_{(l)} b_{{\rm e} (l)}^2/\rho_{(\alpha)}b_{{\rm e}(\alpha)}^2) \Delta L_{(l)}$. 
The last condition establishes the requirement that the number of the blobs be the same (equal to $N_{CG}$). 
Identifying ${\bar{N}_{(l)}}$ and matching $\Delta L_{\rm (\gamma)}$ requires chemistry-specific information, 
i.e. $b_{\rm e (\gamma)}^2$. Thus conventional microscopic simulations of small samples of melts 
with moderate chain-lengths are performed. $R_{b{\rm (\gamma)}}^2$ is calculated and fitted 
with the RHS of Eq.~\ref{eqn:strasbourg}, to extract $b_{\rm e (\gamma)}^2$, $c_{\rm 1}$, and $c_{\rm 2}$. These data provide also an estimate of $\Delta L_{\rm (\gamma)}$ at which 
the ideal random walk approximation is accurate. To perform the reference simulations several techniques are available, including configuration-assembly~\cite{Auhl, Livia} 
and rebridging Monte Carlo algorithms~\cite{Doros}.

Prior to fine-graining, the microscopic length scale is recovered back-transforming (cf. Fig.~\ref{fig:summary}) the coordinate space 
of the blob-based description as ${\bf r} = \Delta L_{\rm (\alpha)} \tilde{\bf{r}}$. In principle $\Delta L_{\rm (\alpha)}$ (equivalently $N_{b{\rm (\alpha)}}$) 
can be large, hampering direct reinsertion of microscopic details into the common blob-based representation. This difficulty can be circumvented~\cite{Guojie2014} through a hierarchical scheme 
where a low-resolution blob-based description undergoes a sequence of fine-graining steps. Each step increases resolution substituting every blob with a pair of smaller ones. 
On these intermediate scales conformations and liquid structure are affected substantially by microscopic features, so that the blob-based 
description becomes chemistry-specific. Nevertheless, fine-graining is feasible~\cite{Guojie2014} considering the blobs as soft spheres~\cite{Vettorel} with simple
interactions parameterized from the reference microscopic simulations. Alternatively, somewhat more complex chemistry-specific blob-based 
models can be developed from integral equation theories~\cite{Guenza2013, Guenza2014, DavidWang}, reducing in the future the need for such calibration data. 
Once the blobs become sufficiently small, microscopic details can be introduced. The general strategy~\cite{PeterKremer} is to reinsert the underlying subchains 
so that they comply, e.g. with the size and location of the COM of the blobs, and relax the reintroduced degrees of freedom.

As a proof-of-principle, we consider here homopolymer melts described with a generic microscopic model~\cite{KG}. 
Linear chains are represented with monomers linked by FENE springs augmented by an angular potential~\cite{Auhl} $U({\theta})=\kappa_{\theta (\gamma)}(1-\cos{\theta_{\rm o}})$, 
where $\theta_{\rm o}$ is the angle between two sequential springs. Nonbonded interactions are captured through a Weeks-Chandler-Andersen (WCA) potential. 
For FENE and WCA interactions the standard parameterization~\cite{KG} is employed. We mimic chemical diversity by varying chain stiffness and/or 
monomer number density in the range $0\le\kappa_{\theta (\gamma)}\le 1.5 $ and $0.60 \le \rho_{\rm (\gamma)} \le 0.85 $, 
respectively. All lengths and energies are expressed in units of $\sigma$ (the characteristic WCA length) and thermal energy, $k_BT$.

\begin{figure}[h]
\includegraphics[width=0.33\textwidth]{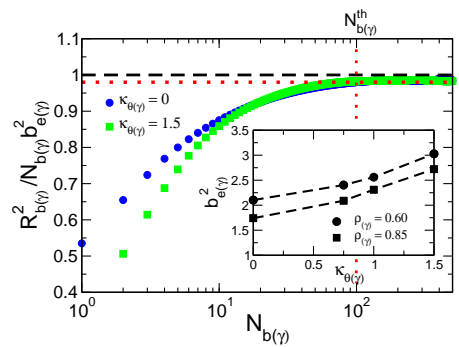}
\caption{Main panel: Mean-square end-to-end distance of subchains, $R_{b{\rm (\gamma)}}^2$, normalized by $N_{b{\rm (\gamma)}}b_{\rm e (\gamma)}^2$ 
as a function of $N_{b{\rm (\gamma)}}$ for two representative values of chain-stiffness parameter, $\kappa_{\theta (\gamma)}$. 
The monomer density is $\rho_{\rm (\gamma)} = 0.85$. The vertical line marks the number of monomers $N_{b{\rm (\gamma)}}^{\rm th} = 100$ 
where deviations from the ideal random walk limit (horizontal dashed black line) are less than $2\%$ (horizontal red dotted line). 
Inset: $b_{\rm e (\gamma)}^2$ as a function of $\kappa_{\theta (\gamma)}$ for $\rho_{\rm (\gamma)} = 0.85$ and $0.60$.}
\label{fig:reference}  
\end{figure}

For the considered $\kappa_{\theta (\gamma)}$ and $\rho_{\rm (\gamma)}$, the required microscopic reference samples contained $400-700$ chains with $N_{\rm (\gamma)} = 500$, at most. 
These moderate sized systems were efficiently equilibrated through a configuration-assembly procedure~\cite{Auhl, Livia} implemented within 
the ESPResSo++ package~\cite{ESPR}. From the reference configurations, $b_{\rm e (\gamma)}^2$ is extracted as previously described (see Figure S1, Tables S1 and S2 in supplemental material~\cite{SuppRef}) and plotted as 
a function of $\kappa_{\theta (\gamma)}$ in the inset of Fig.~\ref{fig:reference}. The main panel of Fig.~\ref{fig:reference} 
presents $R_{b{\rm (\gamma)}}^2/N_{b{\rm (\gamma)}}b_{\rm e (\gamma)}^2$ as a function of $N_{b (\gamma)}$ for $\kappa_{\theta (\gamma)}=0$ and $1.5$. 
Fig.~\ref{fig:reference} demonstrates that for all considered $\kappa_{\theta (\gamma)}$ the deviations from the ideal random walk limit drop below $2\%$ roughly at the same 
threshold $N_{b{\rm (\gamma)}}^{\rm th} = 100$. For practical purposes, for such deviations we accept the ideal random walk statistics as a valid approximation. 
Thus $\Delta L_{\rm (\gamma)}^{\rm th} = b_{\rm e (\gamma)} \sqrt{N_{b{\rm (\gamma)}}^{\rm th}}$ is taken as the smallest possible coarse-graining scale.

All library melts represent the same ''chemical substance'' described by $\kappa_{\theta (l)} = 1.5$ and $\rho_{(l)} = 0.85$.
Creating the library with stiffer homopolymers is advantageous since less monomers per chain are required to achieve
the desired $\bar{N}_{(l)}$, as follows from $N_{\rm (\gamma)} =\bar{N}_{\rm (\gamma)}/\rho_{\rm (\gamma)}^2 b_{\rm e (\gamma)}^6$.
The library samples cover a broad range of invariant degrees of polymerization $7.5\times10^3 \le \bar{N}_{(l)} \le 15\times10^3$ which are representative 
of values in experiments. The samples were generated employing the hierarchical scheme~\cite{Guojie2014} mentioned before. Since the 
universal long-wavelength description is not available at this stage, the melt configuration for fine-graining is obtained~\cite{Guojie2014} from direct 
Monte Carlo equilibration of a chemistry-specific blob-based model. The hierarchical strategy allows the preparation of large 
library melts, e.g. for $\bar{N}_{(l)} = 7.5\times10^3$ and $15\times10^3$ the samples contain $n_{(l)} = 1000$ chains with $N_{(l)} = 500$ and $1000$ monomers, 
respectively.

\begin{figure}[h]
\includegraphics[width=0.5\textwidth]{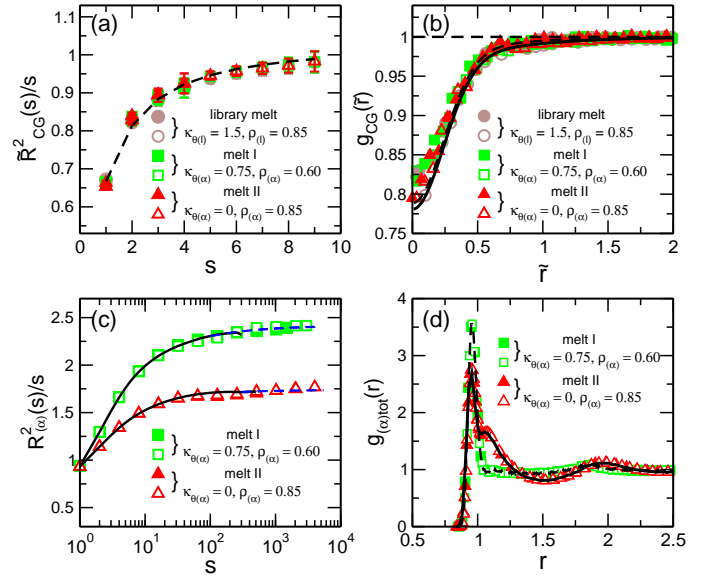}
\caption{(a) Blob-blob internal distance plot, $\tilde{R}_{\rm CG}^2(s)/s$, (in renormalized space) as a function of difference
of their ranking numbers, $s$, in the coarse-grained chain. Solid and open symbols correspond 
to $\bar{N}_{(l)} = 7.5\times10^3$ and $15\times10^3$. Dashed line shows the universal behavior (see  eq.~\ref{eqn:internaldist}). 
(b) Intermolecular pair distribution function of blobs, $g_{CG}(\tilde{r})$, (in renormalized space) calculated for the melts of (a). 
Dashed and solid lines are predictions of eq.~\ref{eqn:guenza} for the two $\bar{N}_{(l)}$. (c) Symbols present the monomer-monomer internal distance plot, $R_{(\alpha)}^2(s)/s$, 
for melts I and II equilibrated by backmapping the universal long-wavelength description. $s$ is the difference of monomer ranking numbers in the microscopic chain. 
Solid lines present $R_{(\alpha)}^2(s)/s$ from reference simulations of shorter melts, extrapolated to larger $s$ (blue dashed lines) substituting into eq.~\ref{eqn:strasbourg}
the fitted $b_{\rm e (\gamma)}^2$, $c_{\rm 1}$, and $c_{\rm 2}$. (d) Total pair distribution function of monomers, $g_{(\alpha) {\rm tot}}(r)$, for the melts of (c).}
\label{fig:results}
\end{figure}

In the library melts, the universal description is reproduced with high accuracy already on the threshold scale, $\Delta L_{(l)}^{\rm th} \simeq 16.5$. 
As an illustration, Fig.~\ref{fig:results}a presents for two melts with $\bar{N}_{(l)} = 7.5\times10^3 $ and $15\times10^3$ the internal 
distance plot $\tilde{R}_{\rm CG}^2(s)/s$. Within error bars the plots (solid and open circles) follow each other, reproducing closely the universal 
behavior of eq.~\ref{eqn:internaldist} (dashed line). The magnitude of deviations is quantified~\cite{SuppRef} in Figure S2. 
The blob-blob intermolecular pair distribution function is presented in Fig.~\ref{fig:results}b (solid and open circles) and is well 
described by the universal form in eq.~\ref{eqn:guenza}.  

Configurations of all other systems can be obtained by fine-graining the blob-based representation characterizing a library melt with the same $\bar{N}_{(l)}$, 
on the scale $\Delta L_{(l)} = 16.5$. As an example, we discuss two chemically different melts 
with $\kappa_{\theta (\alpha)} = 0.75$, $\rho_{\rm (\alpha)} = 0.60$ (melt I) and $\kappa_{\theta (\alpha)} = 0$, $\rho_{\rm (\alpha)} = 0.85$ (melt II). 
From the definition of invariant degree of polymerization and the data on $b_{\rm e (\alpha)}^2$ from Fig.~\ref{fig:reference}, it follows that 
both melts for $N_{\rm (melt\;I)} = 1500$ and $N_{\rm (melt\;II)} = 2000$ map on $\bar{N}_{(l_1)} = 7.5\times10^3$. For $N_{\rm (melt\;I)} = 3000$ and $N_{\rm (melt\;II)} = 4000$ 
both correspond to $\bar{N}_{(l_2)} = 15\times10^3$. Because of scale matching, the universal representation corresponds to $\Delta L_{\rm (melt\;I)} = 26.9$ and $\Delta L_{\rm (melt\;II)} = 26.4$, 
representing an underlying number of monomers $N_{b{\rm (melt\;I)}} = 300$ and $N_{b{\rm (melt\;II)}} = 400$, respectively. The universal description is hierarchically fine-grained~\cite{Guojie2014} 
(cf. Fig.~\ref{fig:summary}) until reaching blob-based descriptions corresponding to $N_{b{\rm (melt\;I)}} = 75$ and $N_{b{\rm (melt\;II)}} = 50$. 
Microscopic details are introduced into these small blobs, employing constrained reinsertion of underlying subchains and gradual elimination of overlaps between monomers~\cite{Auhl, Livia}. 

To demonstrate equilibration of melts obtained from this procedure, Fig.~\ref{fig:results}c presents their internal 
distance plots, $R_{\rm (\alpha)}^2(s)/s$, calculated on microscopic basis. $s$ denotes the difference 
of the ranking numbers of monomers along chain contour and $R_{\rm (\alpha)}^2(s)$ is their mean-square end-to-end distance. 
In all cases $R_{\rm (\alpha)}^2(s)/s$ in the backmapped melts follow closely their counterparts in {\it independent} 
reference simulations (they deviate at most by $2\%$, see~\cite{SuppRef} Figure S3). These observations strongly verify our approach, 
since internal distance plots are extremely sensitive~\cite{Auhl} to the quality of equilibration. Fig.~\ref{fig:results}d compares the total monomer-monomer 
pair distribution functions, $g_{(\alpha) {\rm tot}}(r)$, in the backmapped and reference melts, demonstrating that the local structure is correctly reproduced. 
It is instructive to convert the equilibrated melts I and II to blob-based representations in the renormalized space, 
corresponding to the initial coarse-graining scales $\Delta L_{\rm (melt\;I)} = 26.9$ and $\Delta L_{\rm (melt\;II)} = 26.4$. 
The calculated $\tilde{R}_{\rm CG}^2(s)/s$ and $g_{CG}(\tilde{r})$ are plotted in Fig.~\ref{fig:results}a, b and reproduce the universal behavior of the library melts.

These results confirm that a single melt can serve as a blueprint for configurations of different homopolymers 
described with microscopic resolution. Describing all systems through bead-spring models does not compromise the applicability of the approach to real materials. 
It has been already demonstrated that such models offer microscopic (albeit coarse-grained) description of real polymers when 
appropriately parameterized~\cite{CGPC1998,CGPE2002,CGPS2006}. Moreover, the atomistic description can be recovered by backmapping the bead-spring 
representation through standard schemes~\cite{CGPC1998,CGPS2006}. So far, large samples with experimentally-relevant invariant 
degrees of polymerization, as those in the current work, have been generated only on drastically coarse-grained level~\cite{MuellerSoft}. 
Here these systems are described microscopically with minimum computational costs. For example, equilibrating a sample of melt II 
with $N_{\rm (melt\;II)} = 4000$ at $\bar{N}_{(l)} = 15\times10^3$ ($4\times10^6$ monomers in total) required only six days on $32$ 
processors ($3.0$\;GHz). The concept was elaborated and verified for melts of linear chains but should be straightforward to 
extend to certain cases of non-linear molecules, e.g. branched or concatenated ring-polymers (named~\cite{deGennes} ``Olympic gels''). 
In contrast, the implementation to nonconcatenated ring-polymers is challenging since coarse-graining into blobs relaxes microscopic non-crossability of subchains. 
It is important to explore whether similar strategies are applicable to materials with complex long-wavelength structure, depending on more 
control parameters~\cite{Fredrickson1987, Morse2014}, e.g. microphase-separated block-copolymers.

It is a pleasure to thank Livia Moreira for fruitful discussions and Carlos Marques for carefully reading our manuscript.
The computing time granted by the John von Neumann Institute for Computing (NIC) on the supercomputer JUROPA at J{\"u}lich Supercomputing Centre
(JSC) is gratefully acknowledged.



\begin{thebibliography}{999}
\bibitem{typexplan} Two homopolymer melts are of different type, if their chains are unlike in some respect. Apart from cases of 
dissimilar chemical structure this definition includes differences, e.g. in polymerization degree of the two melts.
\end{thebibliography}
\end{document}